\documentclass[a4paper,12pt]{article}
\usepackage{theorem}
\usepackage{amsfonts}

\theoremstyle{change}
\newtheorem{thm}{Theorem.}[section]
\newtheorem{lem}[thm]{Lemma.}
\newtheorem{cor}[thm]{Corollary.}
\newtheorem{prop}[thm]{Proposition.}
\theorembodyfont{\rmfamily}
\newtheorem{dfn}[thm]{Definition.}
\newtheorem{expl}[thm]{Example.}
\newtheorem{expls}[thm]{Examples.}

\newtheorem{probs}[thm]{Problems.}

\def\Box{\hbox{\raisebox{0.0em}{\rlap{$\sqcap$}}\kern0em%
            \raisebox{-0.0em}{$\sqcup$}} } 
\newenvironment{proof}{{\it Proof. }}{\hfill$\Box$\vspace{0.5cm}}
\def\bepf{\begin{proof}}
\def\epf{\end{proof}}

\def\bary{\begin{array}}
\def\eary{\end{array}}
\def\beq{\begin{equation}} 
\def\eeq{\end{equation}} 
\def\lbeq#1{\begin{equation} \label{#1}} 
\def\gzit#1{{\rm (\ref{#1})}} 	
\def\fct#1{\mathop{\rm #1}}	

\def\Cz{\mathbb{C}}

\def\Ez{\mathbb{E}}

\def\Hz{\mathbb{H}}

\def\Pz{\mathbb{P}}

\def\Rz{\mathbb{R}}

\def\newl{\hfill\break}				
\def \D{\displaystyle}

\def\hbar{h\hspace{-2mm}^-}
\def\kbar{k\hspace{-2mm}^-}
\def\<{\langle} 				
\def\>{\rangle} 				
\def\implies{\quad \Rightarrow \quad}
\def\iff{\quad \Leftrightarrow \quad}
\def\tr{\fct{tr}}
\def\pr{\fct{pr}}
\def\re{\fct{Re}}
\def\im{\fct{Im}}
\def\cov{\fct{cov}}
\def\spec{\fct{Spec}}
\def\half{\frac{1}{2}} 
\def\shalf{\mbox{\small$\frac{1}{2}$\normalsize}}
\def\eps{\varepsilon}
\def\wave{\raisebox{-0.6ex}{\symbol{126}}}

\begin{document}

\vspace*{-2cm}

\begin{center}

{\LARGE \bf Noncommutative analysis and quantum physics} \\
{\LARGE \bf I. States and ensembles}

\vspace{1cm}

\centerline{\sl {\large \bf Arnold Neumaier}}


\centerline{\sl Institut f\"ur Mathematik, Universit\"at Wien}
\centerline{\sl Strudlhofgasse 4, A-1090 Wien, Austria}
\centerline{\sl email: neum@cma.univie.ac.at}
\centerline{\sl WWW: http://solon.cma.univie.ac.at/\wave neum/}

\end{center}

\vspace{0.5cm}

{\small
{\bf Abstract.} 
In this sequence of papers, noncommutative analysis is used to give a 
consistent axiomatic approach to a unified conceptual foundation of 
classical and quantum physics.

The present Part I defines the concepts of observables, states and 
ensembles, clarifies the logical relations and operations for them, 
and shows how they give rise to dynamics and probabilities.

States are identified with maximal consistent sets of weak equalities 
in the algebra of observables (instead of, as usual, with the rays in a
Hilbert space). This leads to a concise foundation of quantum mechanics,
free of undefined terms, separating in a clear way the deterministic 
and the stochastic features of quantum physics. 

The traditional postulates of quantum mechanics are derived from 
well-motivated axiomatic assumptions. No special quantum logic is 
needed to handle the peculiarities of quantum mechanics. Foundational 
problems associated with the measurement process, such as the 
reduction of the state vector, disappear. 

The new interpretation of quantum mechanics contains `elements of 
physical reality' without the need to introduce a classical framework 
with hidden variables. In particular, one may talk about the state of 
the universe without the need of an external observer and without the 
need to assume the existence of multiple universes.

\vfill
{\bf 1991\hspace{.4em} MSC Classification}: primary 81P10

{\bf 1990\hspace{.4em} PACS Classification}: 03.65.Bz

\begin{flushleft}
\end{flushleft}
}

\newpage 
\section{Introduction} \label{intro}

\hfill\parbox[t]{8.8cm}{\footnotesize

{\em ``Look,'' they say, ``here is something new!''
But no, it has all happened before, long before we were born.}

(Good News Bible, Eccl. 1:10)

\bigskip
{\em Do not imagine, any more than I can bring myself to imagine, 
that I should be right in undertaking so great and difficult a task.  
Remembering what I said at first about probability, I will do my best 
to give as probable an explanation as any other -- or rather, more 
probable; and I will first go back to the beginning and try to speak 
of each thing and of all.}

Plato, ca. 367 B.C. \cite{Pla} 
}\nopagebreak

\bigskip
After more than eleven years of long and exciting, often also 
very frustrating investigations into the foundations of physics, 
one of the most foggy regions of the Platonic world, the fog finally
cleared. In the light of the morning sun, the continent of quantum 
physics appears as a well-organized, comprehensible and beautiful part 
of the Platonic world of precise ideas. 

The present paper is the first one of a series of papers that spell out the fruits of my journeys, designed to give a mathematically 
elementary and philosophically consistent foundation of modern 
theoretical physics, presented in the framework of noncommutative 
analysis. It is an attempt to reconsider, from a more modern point of 
view, {\sc Hilbert}'s \cite{Hil} sixth problem, the 
{\em axiomatization of theoretical physics}. 
It is an attempt only since at the present stage 
of development, I have not yet tried to achieve full mathematical 
rigor everywhere. (However, Parts I and II are completely rigorous,
and in later parts it is explicitly mentioned where the 
standard of rigor is relaxed.)

One of the basic premises throughout my years of search was that the 
split between classical physics and quantum physics should be as small 
as possible. This is optimally realized in the set-up proposed here. 
Except in the examples, the formalism never distinguishes between the 
classical and the quantum situation. Thus it can be considered as a 
consequent implementation of {\sc Bohr}'s {\em correspondence 
principle}. This also has didactical advantages for teaching: 
Students can be trained to be acquainted with 
the formalism by means of intuitive, primarily classical examples at 
first. Later, without having to unlearn anything, they can apply 
the same formalism to quantum phenomena.

\bigskip
The present Part I is concerned with giving a concise foundation by 
defining the concepts of observables, states and ensembles, clarifying 
the logical relations and operations for them, and showing how they 
give rise to the traditional postulates of quantum mechanics,
including dynamics and probabilities.

Much of what is done here is common wisdom in quantum mechanics; 
however, the new interpretation slightly shifts the meaning of some
concepts, fixing them in a way such as to forbid certain 
identifications that gave rise to the riddles of quantum philosophy.
Traditional interpretations always assumed without reflection the 
validity of the equation

\centerline{\em
ideal measurement = state of the system = pure prepared state.
}

The new approach differentiates between the three concepts and gives 
each one a distinct meaning. In particular, we shall strictly 
distinguish between states, measurement and preparation. This helps to 
clarify the meaning of these concepts and reduces the danger of 
paradoxical conclusions. 

By identifying states with maximal consistent sets of weak equalities 
in the algebra of observables (instead of, as usual, with the rays in a
Hilbert space), a concise foundation of quantum mechanics is given,
free of undefined terms. The deterministic and the stochastic features 
of quantum physics are separated in a clear way. No special
quantum logic is needed to handle the peculiarities of quantum
mechanics. Foundational problems associated with the 
measurement process, such as the reduction of the state vector,
disappear. 

The new interpretation of quantum mechanics contains
{\em elements of physical reality} in the sense of 
{\sc Einstein, Podolsky \& Rosen} \cite{EinPR}, without the need to 
introduce a classical framework with hidden variables
(cf. my analysis of flaws in the traditional mathematical arguments 
against realism in \cite{Neu0}).
In particular, one may talk about the state of the universe without
the need of an external observer and without the need to assume the
existence of multiple universes. 

To motivate the new conceptual foundation and to place it into context,
I found it useful to embed the formalism into my philosophy of physics,
while strictly separating the mathematics by using the classical
definition-example-theorem-proof exposition style. Though I present my 
view generally without using subjunctive formulations or qualifying 
phrases, I do not claim that this is the only way to understand physics.
Thus I do not attempt to refute any of the alternative interpretations
or any arguments why particular philosophical positions close to the 
one I maintain are not compelling. However, I do attempt to give a vivid
picture of the particular philosophical view that gave me the vision 
to find the new foundation. And I believe that this view is
consistent with the mathematical formalism of quantum mechanics and
accomodates naturally a number of puzzling questions about the nature
of the world.

\bigskip
Section \ref{statements} motivates and introduces the basic concepts 
of statements, selections, and states, and how they are realized in
classical physics and quantum mechanics. This section also spells out
the relation of the new interpretation to the traditional concepts
of states and ideal measurement. Section \ref{properties} then 
discusses the properties of selections and states in more detail, and
in particular shows that there are states that cannot be described by
a single ideal measurement.

In Section \ref{dynamics} we discuss elementary aspects of the 
dynamics of physical systems, as far as relevant for the new 
interpretation. In Section \ref{ensembles}, we look at the way 
ensembles and probabilities arise in the new interpretation, and
in Section \ref{uncertainty} we consider the implications about the
uncertainties inherent in state preparation and measurement.

The Section \ref{phil} concludes the main exposition by relating the 
new interpretation to more general philosophical questions. We discuss
a particular philosophical position, namely that quantum physics
should be regarded to be as deterministic as classical physics, and 
that the stochastic features arise from the impossibility of preparing 
deterministic states, due to the uncertainty principle. We look at the 
borderline between objective existence and subjective judgment, and 
we reconsider {\em the unreasonable effectiveness
of mathematics in the natural sciences} ({\sc Wigner} \cite{Wig}).

\bigskip
Subsequent parts of this sequence of papers will present 
a differential calculus based on Poisson algebras and its 
application to the dynamics of physical systems \cite{Neu2}, 
the calculus of integration \cite{Neu3} and its application to 
equilibrium thermodynamics \cite{Neu4}, 
a relativistic covariant Hamiltonian multiparticle theory, 
and its application to nonequilibrium thermodynamics 
and quantum field theory.

\bigskip
{\bf Acknowledgments.} 
I'd like to thank Hermann Schichl, Tapio Schneider and Karl Svozil 
for discussions that led to improvements of the present text, 
and Waltraud Huyer for pointing out the need to allow nonhermitian 
observables to get Proposition \ref{p1.2}(iv).

\section{Statements and states} \label{statements}

\hfill\parbox[t]{8.8cm}{\footnotesize

{\it How wonderful are the things the {\sc Lord} does! 
All who are delighted with them want to understand them.}
 
(Good News Bible, Psalm 111:2)
}\nopagebreak

\bigskip
All our scientific knowledge is based on past observation, and only 
gives rise to conjectures about the future. Mathematical consistency 
requires that our choices are constrained by some formal laws. When we 
want to predict something, the true answer depends on knowledge we do 
not have. We can calculate at best approximations whose accuracy 
can be estimated using statistical techniques (assuming that the 
quality of our models is good).

This implies that we must distinguish between {\em quantities} 
(formal concepts of what can possibly be measured or calculated) and 
{\em numbers} (the results of measurements and calculations 
themselves); those quantities 
that are constant by the nature of the concept considered behave just
like numbers. Comparison with experiment exclusively concerns
{\it expectations} of quantities, numbers associated to the quantities
in a precise way depending on the {\em preparation} of an experiment
(cf. Sections \ref{ensembles} and \ref{uncertainty}).

\bigskip
In quantum mechanics (see, e.g., {\sc Jammer} \cite{Jam1,Jam2}, 
{\sc Jauch} \cite{Jau}, {\sc von Neumann} \cite{vNeu}, 
{\sc Messiah} \cite{Mes}), observables are identified with certain
elements of the algebra $\Ez$ of bounded linear operators
in a Hilbert space. However, the Hilbert space has no operational
physical interpretation; hence we shall drop it from our main line of
discussion, and resurrect it only to give examples.

On the other hand, the algebra $\Ez$ is essential. Physicists are used 
to calculating with quantities that they may add and multiply 
without restrictions; if the quantities are complex, the complex 
conjugate can also be formed. Thus we take as primitive objects
of our treatment a set $\Ez$ of quantities, such that the sum and the 
product of quantities is again a quantity, and there is an operation 
generalizing complex conjugation. 

Operations on quantities are required to satisfy a few
simple rules; they are called {\bf axioms} since we take them as a
formal starting point without making any further demands on the
nature of the symbols we are using. Our axioms are motivated by the 
wish to be as general as possible while still preserving the ability 
to manipulate quantities in the manner familiar from matrix algebra. 
(Similar axioms for quantities have been proposed, e.g.,
by {\sc Dirac} \cite{Dir} and {\sc Thirring} \cite{Thi4}.)

\begin{dfn} ~

(i) $\Ez$ denotes a set whose elements are called {\bf quantities}.
For any two quantities $f,g\in\Ez$, the {\bf sum} $f+g$, the 
{\bf product} $fg$, and the {\bf conjugate} $f^*$ are also quantities,
and the following axioms (O1)--(O5) are assumed to hold for 
$f,g,h\in\Ez$ and $\alpha\in\Cz$.

(O1)~ $\Cz \subseteq \Ez$, i.e., complex numbers are special quantities,
where addition, multiplication and conjugation have their traditional
meaning. 

(O2)~ $(fg)h=f(gh)$,~~ $\alpha f=f\alpha $,~~ $0f=0$,~~ $1f=f$.

(O3)~ $(f+g)+h=f+(g+h)$,~~ $f(g+h)=fg+fh$,~~ $f+0=f$. 

(O4)~ $f^{* * }=f$,~~ $(fg)^* =g^* f^* $,~~ $(f+g)^* =f^* +g^* $,

(O5)~ $f^* f =0 \implies f =0$.

(ii) We introduce the traditional notation
\[
-f:=(-1)f,~~ f-g:=f+(-g), ~~~[f,g]:=fg-gf,
\]
\[
f^0:=1,~~ f^l:=f^{l-1}f~~~ (l=1,2,\dots ),
\]
\[
\re f = \half(f+f^*),~~~\im f = \frac{1}{2i}(f-f^*).
\]

(iii) A quantity $f\in\Ez$ is called {\bf Hermitian} if $f^*=f$,
and {\bf normal} if $[f,f^*]=0$. We refer to normal quantities as 
{\bf observables}, and we denote by 
\[
  A _{obs} = \{ f \in A \mid [f,f^*]=0 \}
\]
the {\bf observable part} of a subset $A$ of $\Ez$.

\end{dfn}

(Note that every Hermitian quantity is normal and hence an observable. 
But our definition also allows certain nonhermitian observables.)

We shall see that, for the general, qualitative aspects of the theory
there is no need to know any details of how to actually perform 
calculations with quantities; this is only needed if one wants to 
calculate specific properties for specific systems. In this respect, 
the situation is quite similar to the traditional axiomatic treatment of
real numbers: The axioms specify the permitted ways to handle formulas 
involving these numbers; and this is enough to derive calculus, say,
without the need to specify either what real numbers {\em are} or 
algorithmic rules for addition, multiplication and division. Of course,
the latter are needed when one wants to do specific calculations but not
while one tries to get insight into a problem. And as the development
of pocket calculators has shown, the capacity for 
understanding theory and that for knowing the best ways of calculation 
need not even reside in the same person.

Note that we assume commutativity only between numbers and quantities.
However, commutativity of the addition is a consequence of our other
assumptions:

\begin{prop}
For all quantities  $f$, $g$, $h\in \Ez$,
\[
(f+g)h=fh+gh,~~f-f=0,~~ f+g=g+f.
\]
\end{prop}

\bepf
The right distributive law follows from
\[
\begin{array}{lll}
(f+g)h&=&((f+g)h)^{* *}=(h^* (f+g)^* )^* =(h^* (f^* +g^* ))^* \\
&=&(h^* f^* +h^* g^* )^* =(h^* f^* )^* +(h^* g^* )^* \\
&=&f^{* * }h^{* * }+g^{* * }h^{* * }=fh+gh.
\end{array}
\]
It implies $f-f=1f-1f=(1-1)f=0f=0$. From this, we may deduce that 
addition is commutative, as follows. The quantity $h:=-f+g$
satisfies
\[
-h=(-1)((-1)f+g)=(-1)(-1)f+(-1)g=f-g, 
\]
and we have
\[
f+g=f+(h-h)+g=(f+h)+(-h+g)=(f-f+g)+(f-g+g)=g+f. 
\]
\epf

Thus, in conventional terminology (see, e.g., {\sc Rickart} \cite{Ric}),
$\Ez$ is a {\bf nondegenerate *-algebra with unity}, but not 
necessarily with a commutative multiplication. As the example
$\Ez=\Cz^{n\times n}$ (with complex numbers identified with the
scalar multiples of the identity matrix) shows, $\Ez$ may have zero
divisors, and not every nonzero quantity need have an inverse.
Therefore, in the manipulation of formulas, precisely the same 
precautions must be taken as in ordinary matrix algebra.

\bigskip
A {\em statement} is the assertion of some pieces of 
information available about a system in a given {\em state} or set of 
states. We base our new interpretation on the assumption that the
elementary pieces of information are assertions of weak equality 
$f \approx g$ between quantities $f,g\in \Ez$. 

The {\em logical structure} of the theory is defined by axioms for 
valid inference; these state how valid statements may be 
combined to give further valid statements. Axioms (R1)--(R3) below
express reflexivity, symmetry and transitivity, (R4) 
says that we may add arbitrary quantities to both sides of a
weak equality, and (R5) says that we may multiply weak equalities 
{\em from the left} by arbitrary quantities. On the other hand,
(R6) restricts the addition of weak equalities to those where all
four terms are {\em mutually commuting observables}. This is an 
expression of {\sc von Neumann'}s \cite{vNeu}
observations that the sum of two observables $f,g$ has a natural 
interpretation in terms of $f$ and $g$ only when these and their
conjugates all commute with each other. 
Thus weak equality is indeed a weaker concept than standard equality.
 
\begin{dfn}~

(i) A {\bf statement} is a relation $\approx$ on $\Ez$. We say that 
$f,g\in\Ez$ are {\bf weakly equal} if $f \approx g$, and that
$f\in\Ez$ {\bf vanishes} if $f \approx 0$. In particular, a single
weak equality is considered to be a statement.

(ii) A statement $\approx$ is {\bf logically closed} if, for all 
$f,g,h\in\Ez$,

(R1)~$f\approx f$, 

(R2)~$f \approx g \implies g \approx f$,

(R3)~$f \approx g,~g\approx h \implies f \approx h$,

(R4)~$f \approx g \implies f+h \approx g+h$, 

(R5)~$f \approx g \implies hf \approx hg$,

and, for all $f,g,f',g'\in\Ez_{obs}$ commuting with each other and 
with their conjugates,

(R6)~$f \approx g,~~ f' \approx g' \implies f+f' \approx g+g'$.

(Note that $f\approx g$ usually does {\em not} imply $f^*\approx g^*$!)

(iii) The {\bf logical closure} of a family of statements $\approx_l$
($l\in L$) is the logically closed statement $\approx$ with the fewest
weak equalities satisfying 
\[
f\approx_l g \implies f \approx g
\]
for all $l$ and all $f,g\in\Ez$. We say that any subset of weak 
equalities in the logical closure can be {\bf inferred} from the
statements $\approx_l$ ($l\in L$). The statements $\approx_l$ 
($l\in L$) are called {\bf consistent} if $1 \approx 0$ cannot be 
inferred from them.

\end{dfn}

Traditional quantum logic ({\sc Birkhoff \& von Neumann} \cite{BirN}, 
see also {\sc Svozil} \cite{Svo}) can be regarded as the theory of
weak equalities $e\approx0$ or $e\approx1$ for orthogonal projectors 
$e$. (We shall use these projectors in Section \ref{ensembles} for 
counting {\em events}.)
However, there is no intrinsic reason in the quantum mechanical 
formalism why only these statements should be admissible. 

With our definition, there is no special need for a quantum logic.
The only logic to be used is the classical logic for handling 
inferences about statements, with the rules (R1)--(R6) and standard 
logical operations and quantors. This is satisfying since, indeed, 
classical logic is used in practice to handle virtually all 
applications of quantum mechanics.

\begin{expls}\label{ex2.4}~

(i) {\bf Classical physics.}
Classical physics happens in a set $\Omega$ called the
{\em phase space}, and $\Ez$ is an algebra of bounded functions 
$f: \Omega \rightarrow \Cz$.

If $B$ is an open subset of $\Rz^m$ and $p$ is a vector of commuting
Hermitian observables $p_j$ ($j=1,\dots,m$), we denote by 
$p \approx\in B$ (``$p$ is {\bf weakly in} $B$'') the statement that 
all quantities $F(p)\in \Ez$ with bounded $F:\Rz^m\to \Cz$ and support 
disjoint from $B$ vanish. (If $B$ is not open, $p \approx\in B$ is 
taken to mean $p \approx\in B'$ for all open $B'$ containing $B$.)

In the particular case where $B$ is the open ball with center 
$k\in\Rz^n$ and radius $\eps>0$, we denote this statement by 
$\|p-k\|\approx<\eps$ 
(``$\|p-k\|$ is {\bf weakly smaller} than $\eps$'').
It represents the assertion that a measurement of $p$ would give with 
certainty a value that deviates from $k$ by less than $\eps$.

(ii) {\bf Nonrelativistic quantum mechanics.}
In nonrelativistic quantum mechanics, $\Ez$ is the algebra of bounded 
linear operators in the Hilbert space $L^2(\Omega)$, where $\Omega$ is
the direct product of $\Rz^n$ and a finite set $S$ that takes care of
spin, color, and similar indices. (In contrast to the classical case,
$\Omega$ is only `half' of phase space!)
The statements $p \approx\in B$ and $\|p-k\|\approx<\eps$ have 
precisely the same definition and interpretation as in the classical 
case.

(iii) One of the nontrivial traditional postulates of quantum mechanics,
that the possible values an observable $f$ may take are the 
elements of the spectrum $\spec f$ of $f$, is in the new 
interpretation a simple consequence of the trivial axioms (R1)--(R5);
Let $f\in \Ez$, $\lambda\in\Cz$. 

{\em If the weak equality $f \approx \lambda$ can be deduced from a 
consistent statement then $\lambda \in \spec f$. }

Indeed, if this is not the case
then $g:=(\lambda-f)^{-1}$ exists. By (R5), $f \approx \lambda$
implies $gf \approx g\lambda$, hence 
$1=g(\lambda-f)=g\lambda -gf \approx 0$ by (R4), contradiction.

Note that this holds both in classical physics and in quantum 
mechanics (and more generally whenever $\Ez$ is a Banach *-algebra
\cite{Ric}).

\end{expls}

\bigskip
Because of (R4) we may restrict attention to weak equalities where 
one side is zero. We therefore define the concept of a {\em selection} 
that serves to characterize the set of vanishing quantities that can be inferred from a given set of statements. A selection can be thought of 
as containing all information deducible from partial knowledge about 
a system in a given state.

\begin{dfn}~

(i) A {\bf selection} is a nonempty subset $\Pi$ of $\Ez$ such that

(S1) ~$f \in \Ez,~g \in \Pi \implies  fg \in \Pi $,

(S2) ~$f,g \in \Pi _{obs},~ [f,g]=[f,g^*]=0 \implies  f + g \in \Pi$.

A selection is called {\bf ideal} if, in place of (S2), the stronger 
statement

(S2a)~$f,g \in \Pi \implies  f + g \in \Pi$, 

holds, and {\bf valid} if $1 \not\in \Pi$. We denote the set of 
selections by $\Pz$. 

(ii) A {\bf state} is a maximal valid selection, i.e., a valid
selection $\Sigma$ such that
\[
  \Sigma \subseteq \Pi \in \Pz,~ \Pi \not= \Sigma \implies  1 \in \Pi.
\]

(iii) A set of selections is called {\bf consistent} if their union is 
contained in a valid selection. Two selections $\Pi$,  $\Pi'$ are 
called {\bf orthogonal} (and we write $\Pi \perp \Pi'$) if 
\[
f \in \Pi,~g\in \Pi' \implies fg^*=0.
\]
An inconsistent pair of orthogonal selections is called an
{\bf alternative}.

(iv) A statement $\approx$ is called {\bf true} in the state $\Sigma$ 
if 
\[
f-g \in \Sigma ~~~\mbox{for all $f,g\in\Ez$ with $f\approx g$}, 
\]
and {\bf false} otherwise.

\end{dfn}

\begin{prop} \label{p1.1} ~

(i) The set of $f \in \Ez$ for which $f \approx 0$ can be inferred from
a given family of statements is a selection.

(ii) If $\Pi$ is a selection then the relation $\approx$ defined by
\[
f \approx g \iff f-g \in \Pi
\]
is a logically closed statement.

\end{prop}

\bepf
This follows directly from the definitions.
\epf

A state contains all possible information about a system, as far as it 
is accessible within the framework of the theory. It asserts 
a maximal set $\Sigma$ of vanishing objects, hence of weak equalities, 
that does not yet contain invalid information that would allow to 
conclude the equation $1 \approx 0$, i.e., $1 \in \Sigma$. 
By maximality, adding any additional information, i.e., a statement 
$f \approx g$ with $f-g \not\in \Sigma$, would allow to deduce the
invalid statement $1 \approx 0$.

In traditional terminology, a maximal valid ideal selection is a 
maximal left ideal of $\Ez$. If $\Ez$ is the algebra of bounded linear 
operators in a separable Hilbert space then every set of the form
\[
  \Pi _\psi = \{ f \in \Ez \mid f \psi = 0 \}
\]
for some vector $\psi \not= 0$ is a maximal left ideal. 
Conversely, all closed maximal left ideals have this form. 
Clearly $\Pi _\psi$ depends only
on the ray $\Cz \psi$ spanned by $\psi$. Thus, in this case, the 
closed ideal selections that are maximal among the valid ones are in 
one-to-one correspondence with the rays in the Hilbert space, i.e., 
with the traditional quantum mechanical pure states. 

According to established quantum mechanical thinking (as codified for
example in {\sc von Neumann} \cite{vNeu}), an ideal measurement defines
a quantum mechanical pure state. It corresponds to a set of weak 
equalities (assertions true in this `pure state'), maximal with respect 
to the restriction that it can (in principle) be verified by an
{\em instantaneous} experiment. 
(See, e.g., {\sc Wigner} \cite[pp.284-288]{Wig2}
for details on the instantaneous approximations involved in ideal
measurements.)

However, as we shall see soon, our concept of a state is essentially 
different from the traditional pure state of quantum mechanics. 
This has very interesting consequences for the interpretation of 
quantum mechanics. In particular, it seems now conceivable that the
simultaneous assertion of precise values for position and momentum 
are consistent (cf. Problem \ref{p3.5}(iv) below).

This is the decisive difference. In the traditional interpretations, 
closed maximal left ideals (rays) are associated with pure states 
obtained by a mysterious process called an `ideal measurement', 
supposedly achieved by `state reduction' through contact with a 
classical measuring apparatus (ill-defined since `classical' has no
meaning in the formalism). And there are severe interpretational 
problems with `superpositions of pure states' that cannot be given
a classical meaning.

In the present interpretation, measurements no 
longer figure in the conceptual basis (though they can be idealized
as being represented by certain ideal selections), and
states are not related to measurement but to the maximal possible 
information that can be asserted without contradiction, namely that 
certain quantities are weakly equal. 

Therefore, the relation of actual experimental measurements to 
ideal measurements is no longer 
a matter obscuring the foundations of physics, but a thermodynamical 
question concerning the interaction of a system with a macroscopic, 
dissipation-producing measuring apparatus (cf., e.g., 
{\sc Davies} \cite{Dav}, {\sc Busch} et al. \cite{BusGL},
{\sc Zurek} \cite{Zur}, {\sc Joos \& Zeh} \cite{JooZ},
{\sc Ghirardi, Rimini \& Weber} \cite{GhiRW}).

The new interpretation has no longer a place for mysteries since all 
concepts used in the interpretation have precise definitions. 
In particular, there is no state reduction, except approximately,
as far as thermodynamical arguments apply. And what was before a
`superposition of pure states' is now simply a particular set of
valid weak equations, without any spooky associations.

\begin{prop} \label{p1.2} Let $\Ez$ be an arbitrary *-algebra.

(i) The only invalid selection is $\Pi = \Ez$.

(ii) Every set of the form 
\[
  \Pi = \Ez g = \{ fg \mid f \in \Ez \},
\]
is an ideal selection.

(iii) If $fg^*=0$ then $\Ez f$ and $\Ez g$ are orthogonal selections.
In particular, if $e^2=e=e^*$ then  $\Ez e$ and $\Ez (1-e)$ form an
alternative.

(iv) If $\Ez$ is commutative, every selection is ideal.
\end{prop}

\bepf
(i) If $\Pi$ is invalid then $1 \in \Pi$ and by (S1) every $f \in \Ez$
is in $\Pi$. Hence $\Pi = \Ez$. 

(ii) and (iii) are straightforward. 

(iv) Since $\Ez$ is commutative, every quantity is normal, hence
$\Pi_{obs}=\Pi$. Thus, since any two quantities commute, (S2) implies 
(S2a), i.e., $\Pi$ is ideal.     
\epf

\begin{expl}\label{exclass}
({\bf Classical physics})

For a set $\Omega$, consider the algebra $\Ez$ of
bounded functions $f: \Omega \rightarrow \Cz$. It is easy to see that 
an ideal selection $\Pi$ is characterized by the set 
\[
  \hat{\Pi} = \{ \omega \in \Omega \mid f ( \omega ) = 0 
  \mbox{ for all } f \in \Pi \}
\]
of points annihilating $\Pi$. Indeed, given $\hat{\Pi}$, the ideal 
selection $\Pi$ can be reconstructed from this set as
$\Pi = \hat{\hat \Pi}$, where, for $W \subseteq \Omega$,
\[
  \hat{W} = \{ f \in \Ez \mid f(\omega) =0 
  \mbox{ for all } \omega \in  W \}.
\]
Conversely, any $\hat W$ is an ideal selection.
The maximal ideals are obtained for $W=\{\omega\}$,
\[
\Pi _\omega =  \{ f \in \Ez \mid f(\omega) =0 \}
\]
with 
$\omega \in \Omega$, in one-to-one correspondence with the points of 
$\Omega$. By Proposition \ref{p1.2}, every selection $\Pi$ is ideal, 
and the states are just the maximal valid ideal selections. Thus the 
set $\Omega$ can be identified with the set of possible states, a 
selection $\Pi$ is equivalent to finding out that the system 
measured is in one of the states $\omega \in \hat{\Pi}$, and an ideal 
measurement is equivalent to finding the precise state of the system. 
This is typical for classical physics, based on a 
{\em commutative} *-algebra $\Ez$.
\end{expl}

\begin{expl} {\bf (Discretized quantum physics)}\label{exquant}

In the algebra $\Ez = \Cz ^{n \times n}$ of $n \times n$-matrices with
complex entries, the ideal selections $\Pi$ are characterized by their 
row space
\[
  \hat{\Pi} = \{ \phi ^* f \mid \phi \in \Cz ^n, f \in\Pi \}.
\]
Indeed, given $\hat{\Pi}$, the ideal selection $\Pi$ can be 
reconstructed from this vector space as $\Pi = \hat{\hat \Pi}$,
where, for vector spaces $W$ of row vectors of length $n$,
\[
  \hat{W} = \{ f \in \Cz ^{n \times n} \mid \phi ^* f \in W
  \mbox{ for all } \phi \in \Cz ^n \}.
\]
Conversely, any $\hat W$ is an ideal selection.
The maximal ideal selections are obtained by choosing for $W$ a 
hyperplane, 
\[
  W_ \psi = \{ \phi ^* \mid \phi\in\Cz^n,\phi ^* \psi =0 \}
\]
for some nonzero $\psi \in \Cz ^n$. Hence they have the form
\[
  \Pi = \hat{W}_\psi = \{f\in\Cz^{n\times n} \mid f\psi=0\}.
\]
Since $\hat{W} _\psi$ does not change when $\psi$ is multiplied by a 
nonzero number, the maximal valid ideal selections (and hence the ideal 
measurements) are in one-to-one correspondence with the rays 
$\Cz \psi$ with $\psi \in \Cz ^n \backslash \{ 0 \}$. However, as we 
shall see in Section \ref{properties}, 
there are many selections that are not ideal, and the
maximal ideal selections are no longer states. This is typical for
quantum physics (though there one generally has a Hilbert space
in place of $\Cz ^n$, and topological considerations modify the
results a little.)
\end{expl}

\bigskip
\section{Properties of selections and states} \label{properties}

\hfill\parbox[t]{8.8cm}{\footnotesize

{\em There are more things in heaven and earth, Horatio,
     than are dreamt of in your philosophy.}

W. Shakespeare, 1602 A.D. \cite{Sha} 
}\nopagebreak

\bigskip
To prove in general the existence of states, we need Zorn's Lemma 
(see, e.g., {\sc Kelley} \cite{Kel}), a well-known consequence of the 
axiom of choice. 

\begin{lem} {\bf (Zorn)}

Let $\cal{F}$ be a family of sets with the {\bf chain property},

(C) The union of every subfamily of $\cal{F}$, linearly ordered 
by inclusion, is contained in some element of $\cal{F}$.

Then $\cal{F}$ contains a maximal element.              
\end{lem}

\begin{thm} \label{t3.2}~

(i) Every valid selection is contained in some state. 

(ii) Every valid ideal selection is contained in a maximal one.

\end{thm}

\bepf                                   
(i) We have to prove that for every valid selection $\Pi$ there is
a maximal valid selection $\Sigma$ containing $\Pi$.
But the form of the axioms (S1) and (S2) implies that the set of
selections has the chain property. Hence the assertion follows
from Zorn's Lemma.

(ii) Similarly, the set of valid ideal selections has the chain 
property, and Zorn's lemma applies.             
\epf

Thus states always exist, though our proof is not constructive.
In fact, Zorn's Lemma is equivalent to the axiom of choice, and hence
cannot be made constructive; so it would be desirable to find a
constructive proof for the case of *-algebras of physical interest. 
On the other hand, this nonconstructiveness might possibly raise 
interesting decidability problems for statements about physics.

We now proceed to get further insight into the 
properties of selections and states.

\begin{prop}~

(i) The intersection of an arbitrary set of selections is again a 
selection.

(ii) The intersection of an arbitrary set of ideal selections is again
an ideal selection. 

(iii) Every subset $A \subseteq \Ez$ is contained in a unique 
smallest selection $\bar{A}$, the intersection of all selections
containing $A$. In particular, writing $\bar{f} = \overline{ \{ f \} }$,
we have
\[
  \bar{0} = \{ 0 \}, ~~~ \bar{1} = \Ez .
\]
(iv) For every subset $A \subseteq \Ez$, the {\bf orthogonal complement}
of $A$, defined by
\lbeq{ecomp}
  A^{\perp} = \{ f \mid f g ^* = 0 \mbox{ for all } g \in A \}
\eeq
is an ideal selection.
\end{prop}

\bepf
Straightforward.             
\epf

\begin{prop}
Two selections $\Pi$ and $\Pi '$ satisfying $1 \in \Pi_{obs} + 
\Pi ' _{obs}$ are inconsistent.
\end{prop}

\bepf
Let $\Pi''$ be a selection containing $\Pi$ and $\Pi'$.
If $1 \in \Pi _{obs} + \Pi ' _{obs}$ then there is a quantity 
$f \in \Pi _{obs}$ such that $1-f \in \Pi ' _{obs}$. Hence 
$f,1-f \in \Pi''_{obs}$. Since $[f,1-f]=[f,1-f^*]=0$, (S2) implies 
$1=f+(1-f) \in \Pi''$. 
\epf

We now show that, in the quantum case,
there are valid selections not covered by an ideal selection.

\begin{thm}\label{t3.5}

Let $\Psi$ be a set of vectors in a Hilbert space $\Hz$ of 
dimension $>1$ such that distinct vectors in $\Psi$ are neither 
parallel nor orthogonal. Then
\[
\Pi_1(\Psi):=\{ \phi\psi^* \mid \phi \in \Hz, \psi\in\Psi \}
\]
is a valid selection in the algebra of bounded linear operators on 
$\Hz$.

\end{thm}

\bepf
Clearly, (S1) is satisfied. Two nonzero operators $f=\phi\psi^*$
and $g=\phi'\psi'^*$ commute iff $\phi' \parallel \phi$ and 
$\psi' \parallel \psi$, and the assumptions on $\Psi$ then imply that
$\psi' = \psi$. Hence (S2) also holds.
\epf

Note that it is easy to choose $\Psi$ in many ways such that the 
subspace spanned by $\Pi_1(\Psi)$ coincides with the space of all
finite rank operators on $\Hz$. The corresponding sets $\Pi_1(\Psi)$ 
are valid selections not contained in a valid ideal selection. 
In particular, by Theorem \ref{t3.2}, this implies that there are 
states not contained in a valid ideal selection (and hence not 
obtainable by an ideal measurement).

\bigskip
In the smallest noncommutative *-algebra $\Ez=\Cz^{2\times 2}$, 
describing a single quantum spin, it is even possible to describe all 
possible selections, and hence all possible states, explicitly. 

\begin{thm}\label{t3.6}~

(i) Every selection $\Pi\neq \bar0,\bar 1$ of $\Ez=\Cz^{2\times 2}$
has the form $\Pi=\Pi_1(\Psi)$ with $\Psi\subseteq\Cz^2$ as in Theorem 
\ref{t3.5}.

(ii) The maximal valid ideal selections are precisely the sets
$\Ez\psi=\Pi_1(\{\psi\})$ with $\psi\in\Cz^2\setminus\{0\}$.

(ii) The alternatives are precisely the pairs $\bar0,\bar 1$ and the 
pairs $\Ez\phi,\Ez\psi$ with nonzero, orthogonal $\phi,\psi\in\Cz^2$.

(iv) The states of $\Ez$ have the form $\Pi=\Pi_1(\Psi)$, where $\Psi$
contains exactly one nonzero vector from each pair of orthogonal 
one-dimensional subspaces of $\Cz^2$.
In particular, no ideal selection is a state.

\end{thm}

\bepf
This is a straightforward consequence of Theorem \ref{t3.5} and 
Example \ref{exquant}.
\epf

The wish to extend this result to more general situations
suggests the following open questions. For $\Ez=\Cz^{2\times 2}$,
the answers to (i) and (ii) are affirmative, and in (iii), no 
additional condition is needed.

\begin{probs}\label{p3.5}~ 

(i) Is every state a union of maximal valid ideal selections?

(ii) Given a state $\Sigma$ and an ideal selection 
$\Pi \not \subseteq \Sigma$, is there always an ideal selection 
$\Pi' \subseteq \Sigma$ not consistent with $\Pi$?

(iii) Let $\Ez$ be a *-algebra (with 1) of linear operators on a 
Hilbert space $\Hz$. If $\Psi$ is a set of vectors from $\Hz$ such 
that distinct vectors in $\Psi$ are neither parallel nor orthogonal, 
which additional conditions must be imposed on $\Psi$ to ensure that
\[
\Pi(\Psi):=\{f \in \Ez \mid f\psi=0 ~\mbox{for some } \psi \in \Psi \}
\]
is a valid selection? 

(iv) Are arbitrarily accurate measurements of position and momentum 
consistent? More precisely, let $p_\mu$ and $q_\mu$ denote the 
(Hermitian) components of the canonical position and momentum vectors, 
with canonical commutation relations 
\[
[p_\mu,p_\nu]=[q_\mu,q_\nu]=0,~~~[q_\mu,p_\nu]=\delta_{\mu\nu}i\hbar.
\] 
Are the statements $\|p-k\| \approx< \eps$ and $\|q-x\| \approx< \eps'$ 
(defined in Example \ref{ex2.4}) consistent whenever $\eps,\eps'>0$?

\end{probs}

\bigskip
\section{Dynamics} \label{dynamics}

\hfill\parbox[t]{6.5cm}{\footnotesize

{\em God does not play dice with the universe.}

Albert Einstein, 1927 A.D. \cite{Ein}
}\nopagebreak

\bigskip
In this section we discuss elementary aspects of the dynamics of 
physical systems, as far as relevant for the new interpretation.
We shall have much more to say about dynamics in later parts of this 
series of papers.

The observations about a physical system change with time. The dynamics 
of a conservative system is described by a fixed (but system-dependent) 
one-parameter family $T_t$ ($t\in\Rz$) of {\bf automorphisms} of the 
*-algebra $\Ez$, i.e., mappings $T_t:\Ez\to\Ez$ satisfying 
(for $f,g \in \Ez$, $\alpha\in\Cz$, $s,t\in\Rz$)
\[ 
T_t(\alpha)=\alpha, ~~~ T_t(f^*)=T_t(f)^*,
\]
\[
T_t(f+g)=T_t(f)+T_t(g), ~~~ T_t(fg)=T_t(f)T_t(g),
\]
\[
T_0(f)=f, ~~~ T_{s+t}(f)=T_s(T_t(f)).
\]
For dissipative systems, a semigroup of mappings $T_t$, $t\geq 0$
with the same properties replaces the group of automorphisms.

In the {\bf Heisenberg picture} of the dynamics, where states are
fixed and quantities change with time,
$f(t):=T_t(f)$ denotes the {\bf Heisenberg quantity} associated with 
$f$ at time $t$. Note that $f(t)$ is uniquely determined by $f(0)=f$. 
Thus {\em the dynamics is deterministic}, independent of whether we 
are in a classical or in a quantum setting.

\begin{expls}
In nonrelativistic mechanics, conservative systems are described by a 
Hermitian quantity $H$, called the {\em Hamiltonian}.

(i) In classical mechanics, a Poisson bracket $\{\cdot,\cdot\}$  
together with $H$ defines the Liouville superoperator $Lf=\{f,H\}$,
and the dynamics is given by 
\[
T_t(f)=e^{tL}(f).
\]

(ii) In quantum mechanics, the dynamics is instead given by
\[
T_t(f)=e^{-tH/i\hbar}fe^{tH/i\hbar}. 
\] 
\end{expls}

Of course, weak equalities valid at some time need not be valid at 
other times. To see what happens, suppose that $f\approx g$ at time 
$t=0$. Then, in the Heisenberg picture, $f(t)\approx g(t)$, i.e.,
$T_t(f) \approx T_t(g)$ at time $t$. Thus, if $\Pi$ is the selection 
of all quantities that can be inferred to vanish at time $t=0$ then
\[
\Pi(t):=\{ T_t(f) \mid f \in\Pi \} 
\]
is the selection of all quantities that can be inferred to vanish at
an arbitrary time $t$. In particular, a state $\Sigma$ at time $t=0$ 
develops into the state 
\[
\Sigma(t):=\{ T_t(f) \mid f \in\Sigma \}
\]
at time $t$. This describes the {\bf Schr\"odinger picture} of the 
state dynamics, where quantities are fixed and states change with time.
Again the dynamics is deterministic.

\bigskip
In a famous paper, {\sc Einstein, Podolsky \& Rosen} \cite{EinPR} 
introduced the following criterion for elements of physical reality:

{\em 
If, without in any way disturbing a system, we can predict with 
certainty (i.e., with probability equal to unity) the value of a 
physical quantity, then there exists an element of physical reality 
corresponding to this physical quantity
}

and postulated that

{\em 
the following requirement for a complete theory seems to be a 
necessary one: every element of the physical reality must have a 
counterpart in the physical theory.
}

Traditionally, elements of physical reality were thought to have to
emerge in a classical framework with hidden variables.
However, to embed quantum mechanics in such a framework is impossible 
under natural hypotheses ({\sc Kochen \& Specker} \cite{KocS}).
 
It is therefore interesting to see that, in the present interpretation, 
each true weak equality is such an element of physical reality. 
In this sense, the new interpretation is a realistic interpretation of 
quantum mechanics.

In particular, one can talk about the state of the universe without
the need of an external observer ({\sc Wigner} \cite{Wig3}) 
and without the need to assume the existence of multiple universes
({\sc Everett} \cite{Eve}). Instead of many worlds, the new 
interpretation suggests that there is a single world, but one with 
more facettes to it than hitherto suspected.

\bigskip
Taking another look at the form of the Schr\"odinger dynamics, we see 
that the vanishing quantities (or equivalently the statements) behave 
just like the particles in an ideal fluid. We may therefore say that 
the Schr\"odinger dynamics describes the {\em flow of truth}
in an objective, deterministic manner. On the other hand,
the Schr\"odinger dynamics is completely silent 
about {\em what} is true. Thus, as in mathematics, where all truth is 
relative to the logical assumptions made (what is considered true at 
the beginning of an argument), in physics truth is relative 
to the initial values assumed (what is considered true at time $t=0$).

In both cases, theory is about what is consistent, and not about what 
is real or true. The formalism enables us only to deduce truth from 
other assumed truths. But what is regarded as true is outside the 
formalism, may be quite subjective and may even turn out to be 
contradictory, depending on the 
acquired personal habits of self-critical judgment. And finding out 
what is `really' true is highly restricted by the quantum mechanical 
uncertainty relations (see Section \ref{uncertainty}); thus different 
experts can only form more or less valid approximations to the real 
truth. This is very much in agreement with what we see in practice.

What we can possibly know as true are the {\em laws} of physics, 
general relationships that appear often enough to see the underlying 
principle (cf. the remarks about induction in the final section). 
But concerning {\em states} (i.e., in practice, boundary conditions) 
we are doomed to idealized, more or less inaccurate 
approximations of reality. {\sc Wigner}'s \cite[p.5]{Wig} expressed
this by saying,
{\em the laws of nature are all conditional statements and they relate 
only to a very small part of our knowledge of the world.}

\bigskip
\section{Ensembles and probability} 
\label{ensembles}

\hfill\parbox[t]{8.8cm}{\footnotesize

{\em We may assume that words are akin to the matter which they 
describe; when they relate to the lasting and permanent and 
intelligible, they ought to be lasting and unalterable, and, as far 
as their nature allows, irrefutable and immovable -- nothing less.  
But when they express only the copy or likeness and not the eternal 
things themselves, they need only be likely and analogous to the real 
words. As being is to becoming, so is truth to belief.}

Plato, ca. 367 A.D. \cite{Pla}

\bigskip
{\em Only love transcends our limitations. In contrast, our predictions
can fail, our communication can fail, and our know\-ledge can fail.
For our knowledge is patchwork, and our predictive power is 
limited. But when perfection comes, all patchwork will disappear.}

St. Paul, ca. 57 A.D. \cite{Pau}
}\nopagebreak

\bigskip
The stochastic nature of quantum mechanics is usually discussed in
terms of {\em probabilities}. 
However, from a strictly logical point of view,
this has the drawback that one gets into conflict with the traditional
foundation of probability theory by {\sc Kolmogorov} \cite{Kol}, 
which does not extend to the noncommutative case. 
Mathematical physicists (see, e.g., 
{\sc Parthasarathy} \cite{Par}, {\sc Meyer} \cite{Mey}) developed a 
far reaching quantum probability calculus based on Hilbert space 
theory. But their approach is highly formal, drawing its motivation 
from analogies to the classical case rather than from the common 
operational meaning.

{\sc Whittle} \cite{Whi} presents a much less known alternative
approach to classical probability theory, equivalent to that 
of Kolmogorov, that treats {\em expectation} as the basic concept and
derives probability from axioms for the expectation. (See the 
discussion in \cite[Section 3.4]{Whi} why, for historical reasons, 
this has remained a minority approach.)  The approach via expectations 
is easy to motivate, leads quickly to interesting results, and extends 
without much trouble to the quantum world, yielding the ensembles
(`mixed states') of traditional quantum physics. 

The axioms we shall require for meaningful expectations are those
trivially satisfied for weighted averages of a finite ensemble of 
observations. While this motivates the form of the axioms and the
name `ensemble' attached to the concept, there is no need at all to 
interpret expectation as an average; this is the case only in 
certain classical situations. 
In general, the expectation of a quantity $f$ is simply a 
value near which, based on the theory, we may expect the measured
value for $f$. At the same time, the standard deviation serves as a 
measure of the amount to which we may expect this nearness to deviate
from exactness. 

\begin{dfn}~

(i) An {\bf ensemble} is a mapping $^-$ that assigns to each quantity 
$f \in \Ez$ its {\bf expectation} $\overline{f}=:\< f\> \in \Cz$ 
such that for all $f,g \in \Ez$, $\alpha \in \Cz$,

(P1)~ $\<1\> =1, ~~\<f^*\>=\<f\>^*,~~ \< f+g\> =\<f\> +\<g\> $, 

(P2)~ $\<\alpha f\> =\alpha\<f\>$, 

(P3)~ If $f \ge 0$ then $\<f\> \ge 0$,

(P4)~ If $f_l\in\Ez,~ f_l \downarrow 0$ then $\inf \<f_l\> = 0$.

Here $f_l \downarrow 0$ means that the $f_l$ converge to $0$ and
$f_{l+1}\leq f_l$ for all $l$.

(ii) The number
\[
\cov(f,g):=\re \<(f-\overline{f})(g-\overline{g})^* \>
\]
is called the {\bf covariance} of $f,g\in\Ez$, and the number 
\[
\sigma(f):=\sqrt{\cov(f,f)}
\]
the {\bf standard deviation} of $f\in\Ez$. 

\end{dfn}

This definition generalizes the expectation axioms of
{\sc Whittle} \cite{Whi} for classical probability theory. 
Note that (P3) ensures that $\sigma(f)$ is a nonnegative real number
that vanishes if $f$ is a constant quantity (i.e., a complex number). 

To avoid technicalities about topology and order relations 
(discussed in a more detailed treatment in Part III \cite{Neu3}), 
we don't use in this section the topological axiom (P4), and
assume that $\Ez$ is either an algebra of functions or the algebra of 
bounded linear operators on some Hilbert space, with partial order 
relation defined by $f\leq g$ iff $g-f$ is Hermitian and nonnegative 
resp. positive semidefinite. In both cases, for all quantities $f,g$, 
\lbeq{eord1}
f^*f\geq 0,~~~ff^*\geq 0,
\eeq
\lbeq{eord2}
g\geq 0 \implies g=g^* \mbox{ and } f^*gf\geq 0.
\eeq

\begin{expls}  \label{ex5.3}~

(i) {\bf Finite probability theory.}
In the commutative algebra $\Ez = \Cz^n$ with pointwise multiplication
and componentwise inequalities, every linear functional on $\Ez$, and 
in particular every ensemble, has the form 
\lbeq{e5.fin}
\<f\>=\sum_{k=1}^n p_k f_k
\eeq
for certain constants $p_k$. The ensemble axioms hold precisely 
when the $p_k$ are nonnegative and add up to one; thus $\<f\>$ 
is a weighted average.

By Example \ref{exclass} (applied to $\Omega=\{1,\dots,n\}$),
the states are precisely the sets 
$\Pi_\omega = \{ f\in\Ez \mid f_\omega =0 \}$
with $\omega \in \Omega$, and the natural probability of a state 
$\Pi_\omega$ in the ensemble defined by \gzit{e5.fin} is $p_\omega$. 
All elementary probability theory with a 
finite number of events can be discussed in this setting.

Note that the probability can be recovered from the expectation
by means of the formula
\lbeq{e5.prob}
\begin{array}{lll}
p_\omega&=&1-\D\sum_{k\neq \omega} p_k
=1-\sup\left\{\D \sum p_k f_k \mid f_\omega =0, f\le 1\right\}\\
&=&1-\sup\left\{\<f\> \ \mid f \in (\Pi_\omega)_{obs}, f\le 1\right\}.
\end{array}
\eeq

(ii) {\bf Quantum mechanical ensembles.}
In the algebra $\Ez$ of bounded linear operators on a Hilbert space 
$\Hz$, traditional quantum mechanics describes a {\bf pure ensemble} 
(traditionally called a `pure state', but this terminology conflicts 
with our new interpretation) by the expectation
\[
\<f\>:=\psi^*f\psi,
\]
where $\psi\in\Hz$ is a unit vector. And quantum thermodynamics 
describes an {\bf equilibrium ensemble} by the expectation
\[
\<f\>:=\tr e^{-S/\kbar}f, 
\]
where $\kbar>0$ is the {\bf Boltzmann constant}, and $S$ is a Hermitian
observable with $\tr e^{-S/\kbar}=1$ called the {\bf entropy} whose 
spectrum is discrete and bounded below.
In both cases, the ensemble axioms are easily verified.

\end{expls}

\begin{prop} \label{p5.2}

For any ensemble,

(i) $f\leq g \implies \<f\> \leq \<g\>$.

(ii) If $f$ is Hermitian then $\bar f = \<f\>$ is real and
\[
\sigma(f)=\sqrt{\<(f-\overline{f})^2 \>}.
\]

\end{prop}

\bepf
(ii) follows directly from (P1), and (i) from (P1) and (P3).
\epf

The interpretation of probability has been surrounded by philosophical
puzzles for a long time; {\sc Fine} \cite{Fin} is probably still the 
best discussion of the problems involved. Our definition generalizes
the classical intuition of probabilities as weights in a weighted 
average and is modeled after the formula \gzit{e5.prob} for finite 
probability theory in Example \ref{ex5.3}(i).

In the special case when a well-defined counting process may be 
associated with the statement whose probability is assessed, our 
exposition supports the conclusion of {\sc Drieschner} \cite[p.73]{Dri},
{\em ``probability is predicted relative frequency''}
(German original: ``Wahrscheinlichkeit ist vorausgesagte relative 
H\"au\-fig\-keit''). More specifically, we assert that, 
{\em for counting events, the probability carries the information of 
expected relative frequency} (see Theorem \ref{t1.5}(v) below). 

To make this precise we need a precise concept of independent events 
that may be counted. To motivate our definition, assume that we look at
times $t_1,\dots,t_N$ for the presence of an event of the sort we want
to count. We introduce observables $e_l$ whose value is the amount
added to the counter at time $t_l$. For correct counting, we need
$e_l\approx 1$ if an event happened at time $t_l$, and $e_l\approx 0$
otherwise; thus $e_l$ should have the two possible values $0$ and $1$ 
only. Since these numbers are precisely the Hermitian idempotents
among the constant quantities, this suggests to identify events with 
general Hermitian idempotent quantities. 

\begin{dfn}~\nopagebreak

(i) A quantity $e \in \Ez$ satisfying
\[
  e^2 = e = e^*
\]
is called an {\bf event}. 
Two events $e,e'$ are {\bf independent} in an ensemble $\< \cdot \>$
if they commute and satisfy
\[
  \< ee' \> = \< e \> \< e' \>.
\]
With any event we associate the ideal selection
\[
  \Pi _e : = \Ez (1-e),
\]
the set of all quantities vanishing as a consequence of $e \approx 1$. 

(ii) In a given ensemble, the number
\[
\pr(\Pi):=1-\sup\{\<f\> \mid f \in\Pi_{obs}, f\le 1\}
\]
is called the {\bf probability} of the selection $\Pi$, and the number 
$\<e\>$ is called the {\bf probability} of the event $e$.
\end{dfn}

\begin{expls}\label{ex1.4}~

(i) {\bf Classical probability theory.} 
In the algebra of bounded complex-valued functions on a set 
$\Omega$, every characteristic function $e = \chi _M$ (with 
$\chi _M (x) =1$ if $x \in M$, $\chi _M (x)=0$ otherwise) is an event, 
and
\[
  \Pi _e = \{ f \in \Ez \mid f(x) =0 \mbox{~for~} x \in M \} ,
\]
\[
   \Pi _e ^\perp = \{ f \in \Ez \mid f(x) = 0 
                   \mbox{~for~} x \not\in M \}.
\]

(ii)  {\bf Quantum probability theory.} 
In the algebra of bounded linear operators on a Hilbert space 
$\Hz$, every unit vector $\psi \in \Hz$ gives rise to an 
{\bf elementary} event $e = \psi\psi ^*$, and
\[
  \Pi _e = \{ f \in \Ez \mid f \psi =0 \},
\]
\[ 
  \Pi _e ^\perp = \{ \phi \psi ^* \mid \phi \in \Hz \}.
\]
(There are also other, nonelementary events.)
\end{expls}

\begin{thm}\label{t1.5}~

(i) For any event $e$, its {\bf negation} $\neg e = 1 - e$ is also an
event, with 
\lbeq{e5.ev}
\Pi _{\neg e} = \Ez e = \Pi _e ^\perp.
\eeq
Moreover, if the weak equality $e \approx \lambda$ 
($\lambda\in\Cz$) can be deduced from a consistent statement 
then $\lambda \in \{0,1\}$. 

(ii) For commuting events $e, e'$, the observables
\[
  e \wedge e' = ee', ~~~ e \vee e' = e + e' - ee'
\]
are also events. Their probabilities satisfy
\[
  \< e \wedge e' \> + \< e \vee e' \> = \< e \> + \< e' \> ,
\]
\[
  \< e \> + \< \neg e \> = 1.
\]
Moreover,
\[
\<e\wedge e'\> = \<e\>\<e'\>~~~\mbox{ for independent events }e, e'.
\]

(iii) For any selection $\Pi$, we have $0\leq \pr(\Pi)\leq 1$. 

(iv) For any event $e$,
\[
  \pr(\Pi_e) = \< e \> , ~~~ \pr( \Pi _e ^\perp) = 1 - \< e \>.
\]
In particular, $0 \le \<e\> \le 1$. 

(v)
For a family of events $e_l$ $(l=1, \ldots , N)$ with
constant probability $\< e_l \> = p$, the {\bf relative frequency}
\[
  q := \frac{1}{N} \D \sum ^N _{l=1} e_l
\]
satisfies
\[
  \< q \> =p;
\]
if the events are independent,
\[
\sigma (q) = \sqrt{ \frac{ p(1-p)}{N}}.
\]
becomes arbitrarily small as $N$ becomes large  
{\bf (weak law of large numbers)}.
\end{thm}

(We remark in passing that, with the operations $\wedge,\vee,\neg$, 
the set of events in any {\em commutative} subalgebra of $\Ez$ 
forms a Boolean algebra; see {\sc Stone} \cite{Sto}.)

\bigskip
\bepf
(i) Clearly $\neg e$ is Hermitian, and 
$(\neg e)^2=(1-e)^2=1-2e+e^2=1-e=\neg e$. Hence $1-e$ is an event.
The left equality in \gzit{e5.ev} holds by definition. Since
$(fe)(g(1-e))^*=f(e-e^2)g^*=0$, we have $\Ez e \subseteq \Pi _e ^\perp$.
But if $f \in \Pi_e ^\perp$ then $f(1-e)^*=0$, hence $f=fe\in \Ez e$,
so that $\Ez e = \Pi _e ^\perp$. Hence \gzit{e5.ev} holds.
Finally, suppose that $e\approx\lambda$. Then $0\approx \lambda-e$ by 
(R4), and by (R5), 
\[
0\approx (1-\lambda-e)(\lambda-e)=\lambda(1-\lambda)-e+e^2
=\lambda-\lambda^2.
\]
If $\lambda \neq 0,1$, we may multiply on the left by the complex number
$(\lambda-\lambda^2)^{-1}$ and find the contradiction $0\approx 1$.
Hence $\lambda \in \{0,1\}$.

(ii) Since $e$ and $e'$ commute, $(ee')^* =e'^* e^* =e'e=ee'$, hence 
$ee'$ is Hermitian; and it is idempotent since 
$(ee')^2=ee'ee'=e^2e'^2=ee'$. 
Finally, $e+e'-ee'=1-(1-e)(1-e')=\neg (\neg e\wedge \neg e')$ 
is an event. The assertions about expectations are immediate.

(iii) Since, by Proposition \ref{p5.2}(i), 
$\<f\>\leq 1$ for all $f\leq 1$, we have $0 \leq \pr(\Pi)$, and since 
$0 \in\Pi$, we have $\pr(\Pi)\leq 1$. 

(iv) Let $\Pi=\Pi_e^\perp$. If $f \in \Pi_{obs}$ then $f=ge$ for some 
$g\in\Ez$, hence $fe=ge^2=ge=f$ and $ef=e^*f^*=(fe)^*=f^*=f$. Now 
$f\leq 1$ implies $1-f\geq 0$, hence $e-f=e^2-efe=e^*(1-f)e\geq 0$, so 
that $\<f\>\leq\<e\>$. Therefore $\pr(\Pi)\geq 1-\<e\>$. Equality holds 
since $f=e\in\Pi_{obs}$ satisfies $1-f=1-e=(1-e)^*(1-e)\geq 0$, 
hence $f\leq 1$. Thus $\pr(\Pi_e^\perp)= 1-\<e\>$. Replacing $e$ by
$\neg e$ and noting (i) gives $\pr(\Pi_e)= \<e\>$.

(v) $\< q \> =p$ follows from
\[
\<q\> =\frac{1}{N}(\<e_1\>+\dots+\<e_N\> )=\frac{1}{N}(p+\dots+p)=p.
\] 
To get the expression for $\sigma(q)$ when the events are independent, 
we first note that
\[
q^2=\frac{1}{N^2}\Big(\sum_je_j\Big)\Big(\sum_ke_k\Big)
=N^{-2}\sum_{j,k}e_je_k.
\]
In the expectation of this sum we get $N^2-N$ contributions of size 
$\< e_j \> \< e_k \>=p^2$ and $N$ contributions of size 
$\< e_j^2 \> = \< e_j \> =p$. Hence 
\[
\< q^2 \> =N^{-2}(Np+(N^2-N)p^2),
\]
\[
\sigma(q)^2= \< (q-p)^2\> =\< q^2\> -2p\< q\> +p^2 =p(1-p)/N.
\]
\epf

Applied to ideal measurements in a pure ensemble, our recipe for 
the probability just gives the classical {\bf squared probability
amplitude} formula:

\begin{cor}
Let $\phi$ be a unit vector in a Hilbert space $\Hz$. In the algebra 
$\Ez$ of bounded linear operators on $\Hz$, the probability 
that a maximal ideal selection of the form
\[ 
\Pi_\phi = \{f\in\Cz^{n\times n} \mid f\phi=0\}
\] 
is valid in a given ensemble is
\[
\pr(\Pi_\phi)=\<\phi\phi^*\>.
\]
In particular, for a pure ensemble described by the unit vector
$\psi\in \Hz$, the probability that $\Pi_\phi$ is valid is
\lbeq{e.sqprob}
\pr(\Pi_\phi)=|\phi^*\psi|^2.
\eeq
\end{cor}

\bepf 
By Example \ref{ex1.4}(iii), $\Pi_\phi=\Pi_e$ with $e=\phi\phi^*$, 
and by Theorem \ref{t1.5}(iv), $\pr(\Pi_\phi)=\<e\>=\<\phi\phi^*\>$. 
In particular, for a pure ensemble described by the unit vector 
$\psi\in \Hz$, we have
$\pr(\Pi_\phi)=\<\phi\phi^*\>=\psi^*\phi\phi^*\psi=|\phi^*\psi|^2$.
\epf

Equation \gzit{e.sqprob} replaces the traditional interpretation of
$|\phi^*\psi|^2$ as the probability that after preparing a pure 
ensemble in `state' $\psi$, an ideal measurement causes a 
`state reduction' to the new pure `state' $\phi$. Note that the new
interpretation of $|\phi^*\psi|^2$ is completely within the formal
framework of the theory and completely independent of the 
measurement process.

\bigskip
\section{Uncertainty} 
\label{uncertainty}

\hfill\parbox[t]{8.8cm}{\footnotesize

{\em The lot is cast into the lap; 
but its every decision is from the {\sc LORD}.}

King Solomon, ca. 1000 B.C. \cite{Sol} 

\bigskip
{\em As the heavens are higher than the earth, so are my ways higher 
than your ways and my thoughts than your thoughts.}

The {\sc LORD}, according to Isaiah, ca. 540 B.C. \cite{Isa} 

\bigskip
{\em Enough, if we adduce probabilities as likely as 
any others; for we must remember that I who am the speaker, and you 
who are the judges, are only mortal men, and we ought to accept the 
tale which is probable and enquire no further.}

Plato, ca. 367 B.C. \cite{Pla} 
}\nopagebreak

\bigskip
The common form and deterministic nature of the dynamics, 
independent of any assumption
of whether the system is classical or quantum, implies that there is
no difference in the causality of classical mechanics and that of 
quantum mechanics. Therefore, the differences between classical 
mechanics and quantum mechanics cannot lie in an assumed intrinsic 
indeterminacy of quantum mechanics contrasted to deterministic 
classical mechanics. 

In the new interpretation of quantum mechanics, no new principle needs 
to be invoked. As in statistical physics, the stochastic nature of 
quantum mechanics can be explained simply by our inability to prepare 
experiments with a sufficient degree of sharpness to pin down the 
state of the system. A `prepared state' is not really a state, in fact 
we usually know little with certainty, and never everything. 
Thus we need to describe the preparation of experiments in a 
stochastic language that permits the discussion of such uncertainties; 
in other words, we shall model prepared experiments by ensembles.

Formally, the essential difference between classical mechanics 
and quantum mechanics in the latter's lack of commutativity.
While in classical mechanics there is in principle no 
limit to the accuracy with which we can approximate a desired state, 
the quantum mechanical uncertainty relation for noncommuting 
observables puts severe limits on the ability to prepare microscopic 
ensembles. Here, {\em preparation} is defined as bringing 
the system into an ensemble such that certain specified weak equalities
hold to an accuracy specified by an explicit standard deviation.

We now discuss the limits of the extent to which this can be done.

\begin{thm} ~\nopagebreak

(i) The {\bf Cauchy--Schwarz inequality}  
\[
|\< fg^* \>|^2 \le \< ff^* \>\< gg^* \>
\]
holds for all $f,g\in\Ez$.

(ii) The {\bf uncertainty relation}
\[
\sigma(f)^2\sigma(g)^2 
\geq |\cov(f,g)|^2+\left|\shalf\<fg^*-gf^*\>\right|^2
\]
holds for all $f,g\in\Ez$.

(iii) For $f,g\in\Ez$, 
\lbeq{ecov1}
\cov(f,g)=\shalf(\sigma(f+g)^2-\sigma(f)^2-\sigma(g)^2),
\eeq
\lbeq{ecov}
|\cov(f,g)| \leq \sigma(f)\sigma(g), 
\eeq
\lbeq{esig}
\sigma(f+g) \leq \sigma(f)+\sigma(g).
\eeq

\end{thm}

\bepf
(i) For arbitrary $\alpha ,\beta\in \Cz$ we have
\[
\begin{array}{ll}
0&\le \<(\alpha f-\beta g)(\alpha f-\beta g )^*\> \\
&=\alpha \alpha ^* \< ff^* \>-\alpha \beta^* \< fg^* \>
-\beta\alpha ^* \< gf^* \>+\beta\beta^* \< gg^* \>\\
&=|\alpha |^2\< ff^* \>-2\re(\alpha \beta^* \< fg^* \>)
+|\beta|^2\< gg^* \>
\end{array}
\]
We now choose $\beta=\< fg^* \>$, and obtain for arbitrary
real $\alpha $ the inequality
\lbeq{f.8}
0\le \alpha ^2\< ff^* \>
-2\alpha |\< fg^* \>|^2+|\< fg^* \>|^2\< gg^* \>.
\eeq
Now $\< gg^* \>\ge 0$ by (P3). If $\< gg^* \>>0$ we can choose 
$\alpha=\< gg^* \>$ and obtain
\[
0\le \< gg^* \>^2\< ff^* \>-\< gg^* \>|\< fg^* \>|^2.
\]
After division by $\< gg^* \>$, we find that (i) holds.
And if $\< gg^* \>=0$ then 
$\< fg^* \>=0$ since otherwise a tiny $\alpha $ produces a negative
right hand side in \gzit{f.8}. Thus (i) also holds in this case.

(ii) Since $(f-\bar f)(g-\bar g)^*-(g-\bar g)^*(f-\bar f)=fg^*-g^*f$,
it is sufficient to prove the uncertainty relation for the case of
quantities $f,g$ whose expectation vanishes. In this case,
\[
(\re \<fg^*\>)^2 +(\im \<fg^*\>)^2 =|\<fg^*\>|^2 \leq 
\< ff^* \>\< gg^* \> = \sigma(f)^2\sigma(g)^2.
\]
The assertion follows since $\re \<fg^*\>=\cov(f,g)$ and
\[
i\im \<fg^*\>=\shalf(\<fg^*\>-\<fg^*\>^*)=\shalf\<fg^*-gf^*\>.
\]

(iii) Again, it is sufficient to consider the case of
quantities $f,g$ whose expectation vanishes. Then
\lbeq{esig1}
\begin{array}{lll}
\sigma(f+g)^2 &=& \<(f+g)(f+g)^*\>
=\<ff^*\>+\<fg^*+gf^*\>+\<gg^*\>\\
&=& \sigma(f)^2+2\cov(f,g)+\sigma(g)^2,
\end{array}
\eeq
and \gzit{ecov1} follows. \gzit{ecov} is an immediate consequence of
(ii), and \gzit{esig} follows easily from \gzit{esig1} and 
\gzit{ecov}.
\epf

In the classical case of commuting Hermitian quantities, the 
uncertainty relation just reduces to the well-known inequality 
\gzit{ecov} of classical statistics. For noncommuting Hermitian 
quantities, the uncertainty relation is stronger. In particular, we may
deduce {\sc Heisenberg}'s \cite{Hei,Rob} uncertainty relation
\[
\sigma(q)\sigma(p)\geq \shalf\hbar
\]
for a pair of {\bf conjugate observables} $p,q$, characterized by 
$[q,p]=i\hbar$ and Hermiticity. Thus {\em no ensemble can be prepared 
where both $p$ and $q$ have arbitrarily small standard deviation}. 
(More general noncommuting Hermitian observables $f,g$ may have 
{\em some} ensembles with $\sigma(f)=\sigma(g)=0$, namely among those 
with $\<fg\>=\<gf\>$.)

We conclude that,
similar to the case discussed by {\sc Schaller \& Svozil} \cite{SchS} 
of a universe generated by a universal discrete computation, in a 
universe containing a conjugate pair of observables, 
an internal observer bound to the laws of this universe
cannot investigate completely the detailed properties of the system. 
However, an external super-observer (viewing this universe as a kind 
of huge computer game and having access to the simulation code) might 
well know everything; at least, the present paper shows such 
a view to be consistent with the mathematics of quantum mechanics.

\bigskip
It is worthwhile to expand the understanding of the uncertainty 
relation by relating it to the restricted additivity of weak equalities
in (R6).

\begin{prop}
Let $p,q$ be Hermitian quantities satisfying $[q,p]=i\hbar$.
Then, for any $k,x\in\Rz$ and any positive $\Delta p,\Delta q \in\Rz$,
\lbeq{e6.unc}
\Big(\frac{p-k}{\Delta p}\Big)^2+\Big(\frac{q-x}{\Delta q}\Big)^2
\geq \frac{\hbar}{\Delta p \Delta q}.
\eeq
\end{prop}
\begin{proof}
The quantities $b=(q-x)/\Delta q$ and $c=(p-k)/\Delta p$ are Hermitian
and satisfy $[b,c]=[q,p]/\Delta q\Delta p=i\kappa$ where 
$\kappa=\hbar/\Delta q\Delta p$. Now the assertion follows from
\[
0\leq (b+ic)^*(b+ic)=b^2+c^2+i[b,c]=b^2+c^2-\kappa.
\]
\end{proof}

Because of Proposition \ref{p5.2}, the left hand side of \gzit{e6.unc} 
cannot have arbitrarily small expectation. Example \ref{ex2.4}(iii) 
implies the even stronger statement that the possible values that a 
measurement of the left hand side of \gzit{e6.unc} can give are the 
odd positive multiples of the right hand side. (Indeed, this is the
spectrum, since $a:=\kappa^{-1/2}(b+ic)$ is a standard annihilation 
operator.) However, since $p$ and $q$ do not commute, it is not 
permitted to deduce from this that the summands 
$(\frac{p-k}{\Delta p})^2$ and 
$(\frac{q-x}{\Delta q})^2$ cannot be both small. Thus we see 
that noncommutativity together with the restricted additivity (R6) of 
weak equalities work together to avoid contradictions between 
the uncertainty principle and possibly precise knowledge of position
and momentum; cf. Problem \ref{p3.5}. 

\bigskip
This points to a significant difference between preparation and 
measurement. The two concepts describe quite different activities: 
In an experiment, preparation always precedes measurement; 
in particular, experiments require a distinguished direction of time 
since the time reverse of an experiment rarely makes sense. 

Moreover, measurement produces {\em new} knowledge (`elements of 
physical reality' in form of weak equations or inequalities) about a 
system, while preparation {\em assumes} statistical knowledge of past 
behavior of the components of a system (`elements of physical 
probability' in form of an ensemble) without any measurement of the
prepared system.

Thus the preparation of systems is provably limited by the uncertainty 
relation for ensembles, while measuring systems seems not to
be limited in the same way. As mentioned in Section \ref{statements},
ideal measurements (corresponding to rays in Hilbert space, hence
mathematically equivalent to pure ensembles and subject 
to the uncertainty relation) are adequate descriptors for (idealized)
measurement processes only if these can be considered instantaneous
({\sc Wigner} \cite[pp.284-288]{Wig2}). 

Therefore, the key to getting more complete information about any 
microscopic system seems to be that {\em one measures properties of 
the system at a number of different times and reconstructs by 
statistical methods the most likely values of the variables of 
interest}, even of conjugate variables like position and momentum. 

An example for this are the 
particle tracks routinely reconstructed in high energy experiments 
(see, e.g., {\sc Bock} et al. \cite{BocGNR}) on the silent assumption 
thay the particles have definite paths at all times. These paths are 
approximated by least squares techniques and provide highly
accurate knowledge about both position and momentum of the particles 
involved.

Thus, for sufficiently small subsystems and sufficiently large 
measuring devices, it seems possible in principle in the new framework 
to find out arbitrarily precisely what has happened, after the fact.
But, since measurements influence the system, this is not possible in 
a way that would allow the precise prediction of the fate of such a 
system in the future, to prepare an ensemble that would violate the
uncertaintly relations.

To which extent one can pin down the `true' state of a system is one of 
the challenges the new interpretation of quantum mechanics given here 
offers. And, considering the possible impact on clarifying the options 
in quantum computing, it might be a challenge with immense practical 
consequences.

One of the basic questions, not yet decided either theoretically or 
experimentally, is whether we can prepare a two-level quantum system 
(a single spin) sharp enough to ensure more than a single bit of 
information (the obvious limit in the classical case).
As we have seen in Theorem \ref{t3.6} a potentially infinite amount of 
information is contained in a quantum spin state. And the prospective 
builders of quantum computers hope that one can exploit the 
quantum properties to beat the classical limitations on computing power.
See, e.g., {\sc Shor} \cite{Sho}, {\sc Braunstein} \cite{Bra}.

However, such a highly informative quantum spin ensemble would have a 
very rugged dependence of the spin on the direction in which it is 
measured, and no one knows how one should prepare such an ensemble. 
Preparing the spin 
to have a fixed value in one particular direction only gives a 
probability distribution for the values in other directions, that 
decreases with the cosine of the angle between a measured direction 
and the prepared direction ({\sc Neumaier} \cite{Neu0}). The question 
is whether a system can be prepared in a way that this probability 
distribution can be sharpened...

\section{Knowledge and physical reality} \label{phil}

\hfill\parbox[t]{8.8cm}{\footnotesize

{\em The man who thinks he knows something does not yet know as he 
ought to know.}

St. Paul, ca. 57 A.D. \cite{Pau2}
}\nopagebreak

\bigskip
Let me discuss here a philosophical statement about the nature of
the world and its relation to what physicists do.

\begin{enumerate}
\item {\em States are}
      (objective existence)
      and change in a deterministic fashion. 
\item {\em Statements are assumed to be known} 
      (deliberate choice using subjective assessment)
      and change in a deterministic fashion. 
\item {\em Ensembles are prepared} 
      (belief calibrated by past observations)
      and give rise to stochastic indeterminacy based on incomplete
      knowledge, that for systems involving a pair of conjugate 
      observables is unavoidable in principle.
\item {\em Measurements are performed} 
      (subjective experience calibrated by training)
      and are stochastically distributed according to the results of
      the quantum mechanical formalism with input from 1.--3.
\end{enumerate}

Expectations are primarily properties {\em not} of reality itself 
but of the ensemble assigned to a system.
The latter depends on our assessment of reality, i.e., on the assumed 
preparation of the device. Thus ensembles only express our {\em
conjectures} (or even prejudice) about reality, and must be brought
into agreement with reality by 
{\em measurement} (finding something out about a system),
{\em pattern recognition} (identifying a substance as hydrogen, say,
by means of a few measurements, and implying then all properties of 
hydrogen for the whole substance), or 
{\em preparation of an experiment} (arranging subsystems whose 
properties are assumed to be known). 
In actual practice, ensembles are always abstractions from 
reality accurate only to a certain extent, and this 
accuracy is assessed by a measurement-assisted subjective 
interpretation of reality. 

From a practical point of view, theory defines what an object is: 
A gas is considered as ideal gas, and a solid as a crystal, if it 
behaves, to our satisfaction, as a model of an ideal gas, or a crystal, 
predicts. And in preparing experiments one uses equipment supposed to
produce a predictable environment; cf. {\sc Wigner}'s \cite[p.5]{Wig} 
statement, 
{\em [In these] machines, the functioning of which he can foresee, 
[...] the physicist creates a situation in which all relevant 
coordinates are known so that the behavior of the machine can be 
predicted.}

Thus, in practice, one never `prepares a state' by what, in the 
traditional foundation of quantum mechanics, is known as an ideal 
measurement; instead, ensembles are prepared by well-informed 
assumptions concerning one's equipment.
(And if experiments don't give the expected results one
usually first checks whether these assumptions were justified!)

Our knowledge about prepared ensembles is obtained only via the observed
behavior in the past in similar situations. This is the operational
meaning of the ensemble -- it is an ensemble chosen on the basis of 
subjective knowledge about {\em past} situations that we hope is 
representative enough to tell us about {\em future} events.

We know that certain materials or machines reliably produce ensembles 
that depend only on variables that are accounted for in our theory and 
that are either fixed or controllable. More precisely, we assume 
that we know this, on the basis of past experience, claims of 
manufacturers, occasional measurements and consistency checks, etc..
Our measure of reliability is a subjective
sense of our satisfaction, or the satisfaction of others whom
we trust, who checked that certain norms are satisfied.
If we are careless or credulous, our subjective knowledge will be far 
off the mark, and the expectations based on it will simply not be 
matched by reality.

This interface between {\em what we understand} and {\em what is}, 
between model and reality, between theory and experiment, between 
calculated expectations and measurements always remains a subjective 
matter. It is ultimately based on trust in measurement devices, 
apparatus specifications, published data, etc., or perhaps rather 
based on trust in the people (including ourselves) responsible for 
them. 

The strength of theoretical physics lies in the fact that it can 
ignore this subjective side by assuming ensembles to be {\em given}, 
which allows one to calculate expectations from well-defined 
assumptions. The weakness of theoretical physics
lies in the impossibility to objectively verify these assumptions;
comparison with reality always rests on trust in subjective aspects
of observation and communication. Science is possible only because
(and in as much as) it is possible to make these subjective aspects
less influential by training people to adhere to high standards of
precision, carefulness and truthfulness.

\bigskip
The new interpretation makes this gap between model and reality very 
explicit by giving precise concepts of states, ensembles and 
expectations. In this way it frees theoretical physics from 
philosophical riddles by a careful cut just at the point where 
objective expectations and their subjective interpretation interact; 
all these riddles are pushed to the subjective side of the cut. 

In this sense, this paper can be viewed as a mathematical commentary 
on the statement of {\sc Margenau} \cite{Mar},
{\em Measurement is ... the contact of reason with nature.}
A three volume work \cite{KraLS,SupKL,Lee} on the foundations of 
measurements gives a comprehensive survey of -- partially 
successful -- attempts to extend the realm of objectivity further by 
axiomatizing the measurement process in classical physics. The 
problems involved for the quantum case are well covered in the reprint 
collection of {\sc Wheeler \& Zurek} \cite{WheZ}. However, one cannot 
avoid making the transition to subjective judgment at {\em some} 
stage, and the setting proposed in the present paper has the great 
advantage of simplicity.

\bigskip
Since this gap between model and reality forms a built-in part of our 
axiomatic treatment, the latter gives a satisfactory account for the 
well-known problem of {\em induction}. Nothing can {\em guarantee} 
that any model is true, even when truth is restricted to `within 
a specified accuracy'. 
(Shall I prove that you live forever? You experienced all your past 
birthdays, without a single exception. Invoking {\sc Ockham}'s razor 
\cite{Ock, HofMC}, {\em frustra fit per plura quod potest fieri per 
pauciora} -- that we should opt for the most economic model 
explaining a regularity, we conclude that this will go on for ever!! 
But, of course, this proves nothing.) 

\bigskip
\hfill\parbox[t]{8.8cm}{\footnotesize

{\em I dreamt that I was in Hell, and that Hell is a place full of all 
those happenings that are improbable but not impossible. 
[...] every time that they have made an induction, the next instance 
falsifies it. This, however, happens only during the first hundred 
years of their damnation. After that, they learn to expect that an 
induction will be falsified, and therefore it is not falsified until 
another century of logical torment has altered their expectation. 
Throughout all eternity surprise continues, but each time at a higher 
logical level.}

B. Russell, 1954 A.D. \cite{Rus}
}\nopagebreak

\bigskip
But any {\em existing} regularity or structure 
in our world can be {\em discovered} by means of induction: diligent 
observation and good theory allows us to {\em formulate} the observed 
regularities as mathematical models. And why can we discover laws of 
nature? Because they can be formulated with few words and formulas -- 
so a limited amount of plausible information allows us to guess 
correctly with almost certainty any law of limited complexity that 
actually exists. This is confirmed by results of {\sc Webb} \cite{Web}
that, at least in applications to machine learning -- the automatic 
discovery of descriptions of massive sets of data from an accessible 
subset of data --, low complexity seems to be an essential element 
in the appropriateness of Ockham's razor.

Thus induction works in physics {\em not} for logical reasons, but 
{\em because nature is so highly structured}. That the latter is the 
case follows from our overwhelming success in describing nature by 
means of concepts and laws of physics. If any method can be 
effective in describing complex structural patterns in nature,
it must use mathematics, the science of exact concepts and 
their relations. Thus {\em the unreasonable effectiveness
of mathematics in the natural sciences} ({\sc Wigner} \cite{Wig}) is
explained by the plain assumption that nature possesses 
highly accurate laws of limited complexity that are universally valid 
and allow one to explain and predict so much about our universe.

\section{Epilogue} \label{epi}

The axiomatic foundation given here of the basic principles underlying 
theoretical physics suggest that, from a formal point of view, the 
differences between classical physics and quantum physics are only 
marginal (though in the quantum case, the lack of commutativity 
requires some care and causes deviations from classical behavior). 
In both cases, everything flows from the same assumptions simply by 
changing the realization of the axioms.

It is remarkable that, in the setting of Poisson algebras described 
and explored in later parts of this series of papers, this remains so 
even as we go deeper into the details of dynamics and thermodynamics.


\end{document}